\documentstyle[12pt,aasms4]{article}

\received{21 January 1998}
\accepted{29 June 1998}

\lefthead{Bock et al.}
\righthead{High Resolution Mid-Infrared Imaging of the Nucleus of NGC 1068}

\begin{document}

\title{High Resolution Mid-Infrared Imaging\\
    of the Nucleus of NGC 1068}

\author{J. J. Bock, K. A. Marsh, M. E. Ressler, and M. W. Werner}
\affil{Jet Propulsion Laboratory, 4800 Oak Grove Drive,
    Pasadena, CA 91109}



\begin{abstract}
We have obtained mid-infrared images of the nucleus of NGC 1068 from
the Hale 5 m telescope at Mt. Palomar with diffraction-limited resolution
and high sensitivity at $\lambda$ = 8.8, 10.3, and 12.5 $\mu$m.
Deconvolved
images show that the infrared emission extends north to south in the inner
2$\arcsec$, consisting of a central peak, a component extending 1$\arcsec$
north of the
central source, a component extending 1$\arcsec$ south of the
central source, and
several smaller structures located 1$\arcsec$ to the
northeast.  The central peak
is extended 0.4$\arcsec$ N-S and unresolved ($\leq$ 0.2$\arcsec$) E-W.  We find
that $50 \pm 5 \%$ of the flux emerges from the
central 0.4$\arcsec$ and that a single
unresolved point source can account for only $27 \pm 5 \%$ of the total flux. 
However, if the central peak arises from optically thick
emission, we estimate that the emitting region has a projected area
$\geq$ 2 pc$^{2}$, and may thus contain a compact source such as a parsec-scale
torus. 
We observe a correspondence
between the northern extension and the northeastern sources appearing on
the mid-infrared images and the [OIII] clouds A-C \& E.  We interpret the
faint optical counterpart to the mid-infrared southern extension as being
due to partial obscuration by the intervening disk of the host galaxy. 
The N-S extension of the mid-infrared emission coincides with one wall of
the conical narrow line region and aligns with the N-S orientation of the
radio jet close to the nucleus.  We interpret the infrared emission as
arising from optically thick dust lining the walls of the low density
cavity formed by the radio jet and heated by radiation from the central
source.
\end{abstract}




%

\section{Introduction}

NGC 1068 represents the closest example of a Seyfert 2 galaxy (D=14 Mpc
assuming H$_{0}$ = 75 km s$^{-1}$ Mpc$^{-1}$, 68 pc/arcsec) and is thus
amenable to study
with high resolution imaging.
The characteristic smooth
optical continuum and broad optical line emission associated with a
Seyfert 1 galaxy can be seen in polarized light (\cite{ant85}),
leading to the interpretation that NGC 1068 harbors an obscured
Seyfert 1 nucleus.  The absence of strong X-ray emission
(\cite{mul92}) indicates that the nucleus is obscured by
a column density of gas in excess of 10$^{24}$ cm$^{-2}$.

The obscuring medium which prevents direct observation of the Seyfert 1
nucleus has an undetermined morphology and may
consist of a parsec-scale torus (\cite{ant93}).  A compact torus
would allow a direct view of the nucleus
when viewed pole-on as in the case of a Seyfert 1 galaxy, but an obscured
view when viewed edge-on as in the case of a Seyfert 2 galaxy,
and would naturally explain the conical
morphology of the narrow-line region (NLR) observed in NGC 1068 and many
other Seyfert galaxies (\cite{mul96}).  Recently,
interferometric radio observations of water masers in the nucleus of
NGC 1068 (Gallimore et al. 1996a; \cite{gre96}) indicate the
presence of molecular material as close as 0.4 pc to the purported nucleus
with Keplerian rotational kinematics.

Mid-infrared observations allow high-resolution, relatively unobscured
imaging of the nucleus of NGC 1068 at the peak of its SED (\cite{roc91}).
Previous mid-infrared observations (\cite{bra93})
resolved approximately half the emission into a spatially extended
component.  Cameron et al. (1993)
proposed that the central source is instead obscured by a large envelope
of clumpy molecular material surrounding the nucleus,
and that the extended mid-infrared emission
originates from molecular clouds exposed to radiation
from the concealed Seyfert 1 nucleus.  The morphology of the material
obscuring the nucleus of NGC 1068 is thus uncertain and may be more complex
than a simple torus.

\section{Observations}

We conducted observations from the Cassegrain focus of the 5 m
Hale telescope on Mt. Palomar on the nights of 1995 December 3-4 using
MIRLIN, a new mid-infrared broadband camera developed at JPL (\cite{res94})
based on the HF-16 128$\times$128 Si:As BIB array fabricated by
Boeing (formerly Rockwell).  The large well depth of $3 \times 10^7$
e$^{-}$/pixel makes the
detector particularly well-suited for use under
the large photon backgrounds encountered in broadband ground-based
mid-infrared astronomy.  Twin filter wheels allow the observer to select a
wide array of filters, including a narrow-band CVF, for observations in the
spectral band 5 $\mu$m - 26 $\mu$m.  We selected 3 bands
within the N-band atmospheric window with moderate bandwidth
($\Delta\lambda/\lambda = 10 \%$), centered at $\lambda$ = 8.8, 10.3,
and 12.5 $\mu$m, for our
observations of NGC 1068.  Although MIRLIN is somewhat more sensitive
observing with a broadband N filter, we chose the narrower bands to probe
the depth of the 10 $\mu$m dust silicate feature.  We used a plate scale
of 0.15$\arcsec$ pixel$^{-1}$ to obtain fully sampled images.

The target was observed in a standard ``chop-nod'' scheme to remove the
foreground emission from the telescope and the atmosphere.  We typically
chopped at 1.25 Hz, coadding several 50 - 100 ms integrations
at each mirror position
to obtain on and off beam exposures A and B.  After 20 s the telescope was
nodded to obtain 2 further
exposures C and D.  The observation cycle was repeated multiple times,
dithering the position of the telescope several arc seconds after each cycle
in order to sample different pixels on the detector.  We obtain a
single background-subtracted image by accurately registering each group
of exposures (A$_{i}$-B$_{i}$)-(C$_{i}$-D$_{i}$) using the bright nucleus
of NGC 1068 and
then coadding,
$\Sigma_{i}$ [(A$_{i}$-B$_{i}$)-(C$_{i}$-D$_{i}$)].

The results reported in this paper are
derived from the data obtained on 3 December with a small chop and nod throw
(both 10$\arcsec$) in order to keep the nucleus on the array at all times and to
realize maximum sensitivity on the compact nucleus.
Observations on 4 December were obtained with a 20$\arcsec$ throw and confirmed
that extended emission around the nucleus is sufficiently compact as to not
appreciably alter the results obtained with the 10$\arcsec$ throw.

During the night of 3 December we observed, in sequential order over 2.5
hours, $\beta$ Andromedae (AM = 1.1), NGC 1068 (1.6), $\beta$ And (1.0),
NGC 1068 (1.35), $\beta$ And (1.0), and $\alpha$ Tauri (1.4) in each of
the 3 wavelength bands.  The measured
flux of NGC 1068 and the reference stars were stable within each observation,
showing $< 5 \%$ fluctuations between exposures due to variations in
atmospheric transparency.  Unfortunately, the appreciable ($20 \%$) change in
flux between the two observations of NGC 1068 and the discrepancy in the
cross-calibration of the reference stars indicate that weather conditions
were not suitable for accurate photometry.

\section{Analysis}

The raw images of NGC 1068, displayed with logarithmic contours in Fig. 1,
represent a signal-to-noise
ratio in total flux in excess of 10$^{3}$ and a peak signal in
excess of 10$^{2}$
times the per pixel noise level, where the noise is estimated from the
off-source data.  The raw images of $\alpha$ Tau
display the
lobed diffraction pattern with 0.5$\arcsec$ width (FWHM) characteristic
of the 5 m Hale telescope and have
a signal-to-noise ratio in total flux greater than 10$^{4}$.

\placefigure{fig1}

\subsection{Deconvolution}

The mid-infrared images are deconvolved using a technique developed by
Richardson \& Marsh (1983) (also see \cite{mar95}) which yields the
most probable image based on the assumptions of Gaussian prior statistics
with positivity in the pre-convolved image and measurement
noise statistics described by a Gaussian with zero mean.  We obtain
qualitatively similar results by instead deconvolving our images using a
maximum likelihood routine.

The deconvolved image depends critically on the quality and stability of
the PSF.  The multiple
observations of $\beta$ And serve to monitor any changes in the
PSF due to atmospheric fluctuations or drifts in the
focus of the telescope.  Because $\alpha$ Tau is matched
in air mass to NGC 1068 during the observations, we selected
it for the PSF in the deconvolution shown in Fig. 1.
We can test the stability of the PSF during the observations by
deconvolving the raw image of NGC 1068 using the PSF from $\beta$ And
instead of $\alpha$ Tau, and
obtain similar results.  We can also
deconvolve the image of $\beta$ And using the
PSF from $\alpha$ Tau,
resulting in a compact image with $\sim$0.2$\arcsec$ width (FWHM) as
shown in Fig. 1.

The deconvolved images of NGC 1068 show a central peak, and resolve
structure elongated roughly N-S.  These data are in general
accord with the deconvolved
images reported by Braatz et al. (1993), with
$\sim$0.7$\arcsec$ resolution (FWHM) after deconvolution, which resolve
emission extended NE-SW.  With higher
angular resolution our data show that the resolved emission closer to
the nucleus extends N-S, as is clearly evident in the raw images
of NGC 1068 (FWHM = 0.5$\arcsec$) and in the inner contours of the Braatz
et al. (1993) image.  The flux of the central region, obtained by
locating a 0.4$\arcsec$ diameter false aperture on the peak, accounts for
$50 \pm 5 \%$ of the measured brightness in these images, consistent
with the result of $54 \%$ from Braatz et al. (1993).  Reconvolving the
deconvolved images with a 0.7$\arcsec$ FWHM Gaussian beam, we obtain
good agreement with the Braatz et al. (1993) data.

Resolved structures located to the northeast of the central peak are
evident at each of the three wavelengths.  Because these structures
have low surface brightness, on order of $2 \%$ of the peak brightness,
their position and flux are more sensitive to
variations in the PSF.  Nevertheless, the
northeastern structures appear with a roughly hook-like geometry in all
three wavelength bands, and are even evident in data (from other nights)
rejected for poor atmospheric stability if deconvolved.  Therefore we
believe that the northeastern structures are not an artifact of
the deconvolution.

In our deconvolved images we find that the central peak
consists of a structure with width
$\sim$0.2$\arcsec$ (FWHM) E-W and
width $\sim$0.4$\arcsec$ (FWHM) N-S.  Because the
width in the E-W direction
is similar to that of the deconvolved image of $\beta$ And
(0.2$\arcsec$ FWHM), we
consider the central peak to still be unresolved E-W.
We can estimate the flux from the unresolved component by assuming that
the deconvolved PSF is a symmetric Gaussian with width 0.2$\arcsec$.
If we extract an unresolved point
source located at the peak, it accounts for $27 \pm 5 \%$ of the total
emission.  The N-S extension of the central peak is better described,
however, by an unresolved 0.4$\arcsec$ uniform line source
accounting for $55 \pm 5 \%$ of the total emission.
Thus at least $73 \%$ of the total flux is resolved N-S in the
deconvolved images.

If the mid-infrared peak contained a significant contribution from a
point source which was partially obscured by dust, one might expect that
the point source would differ spectrally from the surrounding
emission in the 10 $\mu$m silicate feature.  However, the SED of the
extracted point source and
the SED of the residual flux in the
central 0.4$\arcsec$ agree
to within $10 \%$.  This is consistent with
the large column density of obscuring material inferred from X-ray
measurements blocking a direct view of the purported Seyfert 1 nucleus.
We note that the near-infrared point source observed by Thatte et al.
(1997), extrapolating to mid-infrared wavelengths, contributes only
a small fraction ($0.2 - 10 \%$) of the total nuclear flux.

Although weather conditions were not suitable for absolute photometry,
we can probe the depth of the 10 $\mu$m silicate feature.
The measured $\lambda$F$_{\lambda}$(8.5 $\mu$m) /
$\lambda$F$_{\lambda}$(10.3 $\mu$m) and
$\lambda$F$_{\lambda}$(8.5 $\mu$m) / $\lambda$F$_{\lambda}$(12.5 $\mu$m)
flux ratios for the entire nucleus
agree with Roche et al. (1991) to within 10 $\%$ based on the calibration
obtained with $\alpha$ Tau.  We find that the silicate feature is weak
throughout the image, being most pronounced at the mid-infrared peak
where $\lambda$F$_{\lambda}$(10.3 $\mu$m) /
0.5[$\lambda$F$_{\lambda}$(8.5 $\mu$m) +
$\lambda$F$_{\lambda}$(12.5 $\mu$m)] = 0.8.

\subsection{Location of the Nucleus}

In Fig. 2 we compare the deconvolved 12.5 $\mu$m mid-infrared image with
the NLR imaged in the $\lambda$5007 \AA \ emission line of [OIII]
with 0.05$\arcsec$ resolution using the corrected
Hubble Space Telescope (\cite{mac94}).
Because we did not obtain accurate astrometry of the
mid-infrared images, we place the peak of our deconvolved mid-infrared
image at the location of the mid-infrared peak determined by
Braatz et al. (1993).
The [OIII] map shows clumpy emission in a roughly conical
region extended NE whereas the mid-infrared emission is
elongated N-S.  We note the
correspondence between the infrared northern extension and the
[OIII] clouds A-C \& E.  Although the positions of the NE sources are
not well determined, the hook-like NE feature may coincide with the
[OIII] clouds D and F.
Given the limitations of the astrometry,
we cannot discern if the northern extension coincides with the peaks
of the [OIII] clouds A-C \& E, or is displaced
to the west by $\sim$0.15$\arcsec$.
The central peak and southern extension in the mid-infrared emission
do not have distinct counterparts in the [OIII] map.  The northern
extension appears to trace the western wall, and the southern
extension the eastern wall, of the conical NLR
(\cite{pog89}).  With the exception of the weak northeastern sources,
mid-infrared emission is largely absent from the conical region itself.

High resolution radio maps of the nucleus of NGC 1068 resolve 4 distinct
sources and a collimated bipolar jet (Ulvestad, Neff, \& Wilson 1987).
A comparison between the mid-infrared emission and a 5 GHz map of
the radio continuum (Gallimore, Baum, \& O'Dea 1996b) (see Fig. 3),
shows that the N-S extension of the mid-infrared emission
aligns with the N-S orientation of the radio jet close to the nucleus.

\section {Discussion}

The copious infrared nuclear emission of NGC 1068 ($L \sim 10^{11} L_{\odot}$),
originates from thermal dust emission extending out to $\sim$70 pc.
The emission is associated with a gas mass of at least
$M > m_{H} [F_{\nu} / B_{\nu}(T)] D^{2} (10^{23} cm^{-2}) = 10^{4} M_{\odot}$.
Several mechanisms have been
proposed to heat the dust (radiation from a Seyfert 1 nucleus, star
forming regions, shocks).  Tresch-Fienberg et al. (1987) propose that
the northeastern sources might arise from star formation.  In addition
to a lack of evidence for significant star formation near the nucleus
(\cite{cam93}), we find that at higher resolution the
northeastern sources contain only a small fraction ($\sim 10 \%$) of the
total flux.  The nuclear stellar core observed in K-band imaging to
extend out to 2.5$\arcsec$ contributes a minority of the total nuclear
luminosity unless the stars are very young ($<$ 10$^{7}$ years old,
\cite{tha97}).
Capetti, Axon, \& Macchetto (1997a) suggest that shocks
may explain the ionization structure observed in optical line emission.
Unfortunately, the location of these
shock fronts are 4$\arcsec$ from the
nucleus and outside of the central region
we image in the mid-infrared.  Furthermore, shocks fail to provide
sufficient energy to explain the observed mid-infrared flux in the central
region (\cite{cam93}).  Radiation from a Seyfert 1 nucleus
alone is sufficient to supply the necessary luminosity and can heat
dust out to large distances (\cite{bra93}).

The lack of a strong silicate feature might be explained by depletion of
silicates; however, the estimated dust temperature (T $\simeq$ 350 K,
Tresch-Fienberg et al. 1987)
is far below the expected sublimation temperature of
silicate grains (\cite{lao93}).
An arrangement of optically thin dust in emission and absorption could
be invoked to explain the weakness of the silicate feature, but seems
unlikely to hold over the entire image,
including the partially obscured southern extension (see below).  Thus
we conclude the most likely explanation is that the infrared emission
arises from optically thick dust.

\subsection{Northern Extension}

To the north of the nucleus, the mid-infrared emission corresponds to
the peaks (clouds A-F) in the more extended [OIII] emission.
Braatz et al. (1993) proposed that the extended mid-infrared
emission originates from dust associated with the low density NLR.
However, if the mid-infrared emission arises from optically thick dust,
the mid-infrared emitting regions ($A_{V} \geq 20$) must be physically distinct
from the optically thin ($A_{V} \sim 1.5$) NLR.
The mid-infrared emission may originate from dust deep within clumpy
molecular clouds, and the [OIII] emission from the envelope of these
clouds, giving a correspondence between the mid-infrared and [OIII]
emission where the surfaces of the clouds are relatively unobscured.

The beam filling factor of the extended mid-infrared emission may be
calculated by relating the observed surface brightness to that of a
blackbody, $f = \tau f_{A} = I_{\nu} / B_{\nu}(T)$, where $f_{A}$ is
the areal
filling factor and $\tau$ is the optical depth.  Assuming
T = 350 K we obtain $f \simeq 3 \times 10^{-3}$ for the northern and
southern extensions, much smaller than the areal filling factor of the
ionized [OIII] emitting gas, $f_{A} = 0.1$ (\cite{net90}).
Interestingly, the volume filling factor of the extended molecular
gas envelope is $f_{V} < 10^{-3}$ (\cite{rot91}), suggesting
that the mid-infrared emission may correspond to a heated subsection
of the thick ($N \sim 10^{23} cm^{2}$) molecular material surrounding the
nucleus (\cite{bli94}; Planesas, Scoville, \& Myers 1991).

\subsection{Southern Extension}

The southern mid-infrared extension does not appear to have a clear
optical NLR counterpart, although some structure is notable in the
[OIII] map to the south of the nucleus.  A modest quantity of dust
in the intervening disk of the host galaxy may obscure the [OIII]
emission from the southern cone, and indeed a clearer view of the
southern counter-cone may be obtained in polarized near-infrared
images (\cite{pac97}).  We note that the [OIII] to mid-infrared
ratio is 5 times weaker along the southern extension than the northern
extension, giving an obscuration of $A_{V} = 1.7$ if the northern extension
and southern extension are under similar physical conditions.  This
estimate is in general agreement with the column density of gas,
$N_{H} \sim 2 \times 10^{21} cm^{-2}$, estimated by
Macchetto et al. (1994) based on the
reddening of the southern lobe.  We note that asymmetry in the
northern and southern brightness observed in the radio (see Fig. 3)
also holds in the mid-infrared as the northern extension appears
to be approximately twice as luminous as the southern extension.

\subsection{Unresolved Component}

We can estimate the projected area of the emission in the central
region with a simple blackbody argument if the emission is optically
thick, $A = D^{2} F_{\nu} / B_{\nu}(T)$, assuming D = 14 Mpc is the
distance to NGC 1068
and T = 350 K.  Including the entire flux from the central
0.4$\arcsec$, which is well described by a line source, we find that the
emitting area is 4 pc$^{2}$.  Assuming the flux obtained by removing a
point source at the central peak (see section 3.1), the emitting area
is 2 pc$^{2}$.  Thus the mid-infrared peak may harbor a compact source,
such as a parsec-scale dusty torus.  If the emission is obscured by
intervening dust or is optically thin however, the source size will
be underestimated (although the weakness of the silicate feature
throughout the image
suggests that these effects are
small).  If the central source consists of emission from regions of
dust at several different temperatures, partially obscured by cooler
dust (e.g. \cite{pie92}; \cite{efs95}; \cite{hei97}), the spectrum
and effective size of the central source are model dependent.

\subsection{Morphology}

If the mid-infrared emission is indeed powered by radiation from the
central source, then its morphology
denotes regions where radiation from the central source is absorbed by
dust and converted to thermal infrared radiation.  Some of the energy
actually may be reradiated first by hot ($T \sim 1000 K$) dust close to the
central source.  However, the relatively low inferred bolometric
luminosity of the K-band point source observed by Thatte et al. (1997)
in NGC 1068 and the lack of prominent near-infrared emission in the
SED's of Seyfert 1 and 2 galaxies (\cite{roc91}; \cite{mas95})
suggest that the fraction of reradiated energy is modest.  Although a
compact source may be located at the mid-infrared peak,
these data indicate that of the majority
of the radiation from the central source (not including the fraction
which escapes completely) is actually absorbed at distances greater
than 10 pc.  Therefore lines of sight to the nucleus exist which
are intercepted not by a compact torus but by dust located up to
$\sim$70 pc from the nucleus.

As an alternative to a simple torus, we suggest that the morphology of
the extended obscuring medium traced by the mid-infrared emission is
related to the morphology of the radio jets.  We observe that the
mid-infrared emission in the
central 0.4$\arcsec$ is elongated N-S following
the N-S orientation of the radio jet closest to the nucleus.  This
suggests that the mid-infrared emission arises from clouds lining the
walls of a low density cavity heated by radiation from the central
source.  The walls of the low density cavity may be an extension of
the compact structure mapped in H$_{2}$0 maser emission with an opening
angle of $\sim$90$^{\circ}$ (\cite{gre96}).  The resolution of these
deconvolved mid-infrared images is insufficient to distinguish if
the emission in the central 0.4$\arcsec$
has a linear or conical shape.  A compact torus may be responsible for
preventing radiation propagating E-W, explaining the N-S morphology of
the mid-infrared emission.

Further from the nucleus, the northern and southern extensions
of the mid-infrared emission align along a single wall of the conical
ionization region.  Although the filling factor
of the extensions is low, the data cannot distinguish
whether the emission consists of a coherent linear structure extending
N-S of the nucleus which is unresolved E-W, or if it breaks up into
small unresolved clumps.  We note that mid-infrared and [OIII]
emission are absent at the northernmost radio source NE, as would be
expected if the radio jet has swept out a cavity of ionized gas
(Capetti et al. 1997b; Gallimore et al. 1996b).
We also observe that the northern extension terminates approximately
at the location where the radio jet bends eastward, near the radio
component C where the radio jet may be diverted by a giant molecular
cloud (Gallimore et al. 1996b).

\section{Conclusions}

These images resolve the majority of the mid-infrared emission
in the nucleus of NGC 1068 and show that emission from a compact
central source can only account for a fraction of the total flux.
With improved spatial resolution we determine that the resolved
structure extends N-S in the innermost regions,
corresponding to the orientation of the radio jet, with separate
structures located to the northeast of the central source.  We
interpret the mid-infrared emission as arising from optically thick
dust extending out to $\sim$70 pc, associated with a
gas mass $M \geq 10^{4} M_{\odot}$, and heated by radiation from the central
Seyfert 1 nucleus.  In order to explain the observed mid-infrared
morphology we postulate that dust lining the walls of a low-density
cavity is heated by radiation from the central source.

\acknowledgments

The work described in this paper was carried out at the Jet
Propulsion Laboratory, California Institute of Technology, under
an agreement with the National Aeronautics and Space Administration.
We acknowledge the support of the JPL Director's Discretionary Fund
and NASA's Office of Space Science.  Observations at the Palomar
Observatory were made as part of a continuing collaborative agreement
between Palomar Observatory and JPL.  The authors thank the staff at
Mt. Palomar for their assistance during the observations, J. F.
Gallimore for providing the [OIII] and radio images, and David Seib
for providing the focal plane array.

%

\clearpage

\clearpage

\figcaption[mcon.ps]{Panels from left to right show the raw image of
$\alpha$ Tau (PSF), the raw image
of NGC 1068, the deconvolved image of $\beta$ And using the PSF
from $\alpha$ Tau,
and the deconvolved image of NGC 1068 using the PSF from $\alpha$ Tau.
Wavelength bands of observation are 8.8 $\mu$m (top), 10.3
$\mu$m (middle), and 12.5 $\mu$m (bottom).  For the raw images,
contours begin
at the 1$\sigma$ level pixel$^{-1}$ and increase by multiplicative
factors of $\surd$2.
The deconvolved images have 8 contours beginning at 0.8 of the peak
value and decreasing
by multiplicative factors of 2.  The field of view in each panel is
4.8$\arcsec$ $\times$ 4.8$\arcsec$.
\label{fig1}}

\figcaption[oiii.ps]{Comparison of the deconvolved 12.5 $\mu$m image
with the [OIII] $\lambda$5007 \AA \ HST image of Macchetto et al. (1994).
The contour
spacing for the mid-infrared image is as in Fig. 1, and the [OIII]
map is displayed with a logarithmic stretch.
The letters label local maxima in the [OIII] emission as described
by Evans et
al. (1991).  The green circle denotes the location of the
mid-infrared peak
determined by Braatz et al. (1993), the square denotes the location
of the center of UV polarization (determined most recently by
Capetti et al. 1997b), and the cross denotes
the location of the near-IR peak determined by Thatte et al. (1997).
Blue X's denote the location of the 4 radio peaks (S2, S1, C and NE
from south to north) assuming the registration of Gallimore et al.
(1996b).  The water masers located at the S1 radio source trace out a
velocity gradient which can be described as an edge-on Keplerian disk.
The size of each green symbol is equal to the
positional uncertainty quoted by the authors.
\label{fig2}}

\figcaption[radio.ps]{Comparison of the deconvolved 12.5 $\mu$m image
with the 5 GHz map of
Gallimore et al. (1996b).  We register the maps by simply
aligning the mid-infrared peak with S1, consistent to within
the error in registering the two images to the [OIII] map.
\label{fig3}}











\end{document}